\documentclass[useAMS,usenatbib]{mn2e}
\usepackage{amssymb}
\usepackage{graphicx}
\usepackage{hyperref}
\usepackage{color}
\usepackage[none]{hyphenat}

\title[The Location of Sgr A*]{Pinpointing the near-infrared location of Sgr A* by correcting optical distortion in the NACO imager}
\author[Plewa et al.]{
P. M. Plewa$^{1}$\thanks{E-mail: pmplewa@mpe.mpg.de}, 
S. Gillessen$^{1}$,
F. Eisenhauer$^{1}$,
T. Ott$^{1}$,
O. Pfuhl$^{1}$,
E. George$^{1}$,\newauthor
J. Dexter$^{1}$ ,
M. Habibi$^{1}$,
R. Genzel$^{1,2}$,
M. J. Reid$^{3}$,
K. M. Menten$^{4}$
\\
$^{1}$Max-Planck-Institut f\"ur Extraterrestrische Physik, D-85748 Garching, Germany\\
$^{2}$Department of Physics and Astronomy, University of California, Berkeley, Le Conte Hall, Berkeley, CA 94720, USA\\
$^{3}$Harvard-Smithsonian Center for Astrophysics, 60 Garden Street, Cambridge, MA 02138, USA\\
$^{4}$Max-Planck-Institut f\"ur Radioastronomie, Auf dem H\"ugel 69, D-53121 Bonn, Germany\\
}
\begin{document}

\date{4 September 2015}

\pagerange{\pageref{firstpage}--\pageref{lastpage}} \pubyear{\the\year}

\maketitle

\label{firstpage}

\begin{abstract} 
Near-infrared observations of stellar orbits at the Galactic Center provide conclusive evidence for a massive black hole associated with the compact radio source Sgr A*. The astrometric reference frame for these observations is tied to a set of red giant stars, which are also detectable at radio wavelengths through SiO maser emission in their envelopes. We have improved the precision and long-term stability of this reference frame, in which Sgr A* is localized to within a factor $5$ better than previously: ${\sim0.17\rmn{~mas}}$ in position (in 2009) and ${\sim0.07\rmn{~mas~yr}^{-1}}$ in velocity. This improvement is the result of modeling and correcting optical distortion in the VLT/NACO imager to a sub-mas level and including new infrared and radio measurements, which now both span more than a decade in time. A further improvement will follow future observations and facilitate the detection of relativistic orbital effects.
\end{abstract}

\begin{keywords}methods: data analysis - techniques: high angular resolution - astrometry - Galaxy: centre - infrared: stars\end{keywords}

\raggedbottom

\section{Introduction}
\label{sec:introduction}

The observability of the Galactic Center in the near-infrared and its relative proximity at a distance of ${\sim8~\rmn{kpc}}$ allow the study of a galactic nucleus in unparalleled detail \citep[for a review see][]{2010RvMP...82.3121G}. Adaptive optics assisted imaging observations of the Galactic Center with NACO at the VLT or NIRC2 at the Keck observatory routinely achieve a resolution close to the diffraction limit  and a depth limited by source confusion. Since the discovery of stars with large proper motions \citep{1996Natur.383..415E,1998ApJ...509..678G}, the continued monitoring of individual stellar orbits has produced strong evidence for a compact mass of ${\sim4\cdot10^6M_\odot}$ that can be convincingly attributed to a quiescent massive black hole \citep[e.g.][]{2008ApJ...689.1044G,2009ApJ...692.1075G} at the location of the radio source Sgr~A* \citep{1997ApJ...475L.111M,2003ApJ...587..208R,2007ApJ...659..378R}. Increasing the astrometric precision would permit measuring orbits of more stars than presently possible, permit refined estimates of the black hole's mass and distance, and eventually permit the detection of relativistic effects on the orbits of short-period stars, most immediately post-Newtonian effects on the orbit of the star S2 \citep[e.g.][]{1998AcA....48..653J,2000ApJ...542..328F,2001A&A...374...95R,2005ApJ...622..878W,2006ApJ...639L..21Z,2007PASP..119..349N,2008ApJ...674L..25W,2010PhRvD..81f2002M,2010ApJ...720.1303A,2011Msngr.143...16E}.

While the motions of the so-called S-stars in the central $1''$ are measured from images taken with the S13 camera of NACO \citep[${\sim13~\rmn{mas~px}^{-1}}$, see ][]{1998SPIE.3353..508R,1998SPIE.3354..606L}, less frequent observations with the S27 camera (${\sim27~\rmn{mas~px}^{-1}}$) are used to efficiently set up the astrometric reference frame \citep{2008A&A...492..419T}. However, the S27 camera is affected by optical image distortion on a non-negligible level, as for example is the narrow field camera of NIRC2 \citep{2010ApJ...725..331Y}. By systematically altering the relative positions of stars on the detector, this effect becomes a substantial source of uncertainty when positions measured on widely dithered images are combined into a common coordinate system or when they are aligned with astrometric reference sources.

In our earlier studies of stellar orbits at the Galactic Center \citep{2002Natur.419..694S,2003ApJ...596.1015S,2003ApJ...597L.121E,2005ApJ...628..246E,2009ApJ...692.1075G} we used high-order coordinate transformations to mitigate the effect of image distortion and relied primarily on a reference frame based on the assumption that a sufficiently large number of evolved stars in the nuclear cluster show no net motion. The stellar motions in this `cluster rest frame' were then placed in an astrometric coordinate system centred on Sgr~A* by aligning the positions of several SiO maser stars at a certain epoch with their positions as predicted from radio observations. The comparison of the S-stars' positions over several years was still limited by a systematic uncertainty of ${\sim2~\rmn{mas}}$ associated with the definition of the coordinate system, likely because of residual distortion in the S27 camera. Yet the typical uncertainty of a single position is as low as ${\sim0.3~\rmn{mas}}$, limited by residual distortion in the S13 camera, uncertainty in modelling the point spread function (PSF) and ultimately source confusion \citep{2010MNRAS.401.1177F}. The long-term stability of the cluster rest frame is also fundamentally limited to ${\sigma/\sqrt{n}\approx0.07\rmn{~mas\ yr}^{-1}}$ by the intrinsic velocity dispersion of the selected stars in the plane of the sky (${\sigma\approx3\rmn{~mas\ yr}^{-1}}$; \citealt{2008A&A...492..419T}) and the number of available stars (${n\approx2000}$; \citealt{2009ApJ...692.1075G}).

In this paper we construct an alternative astrometric reference frame, which has now become more advantageous, by relating many well-measured stars to the SiO maser stars directly at multiple epochs and then using their motion with respect to Sgr~A* for reference. The details on the construction of this `Sgr~A* rest frame' \citep{2010ApJ...725..331Y} are described in section~\ref{sec:astrometry} and the precision of localizing (radio-)Sgr~A* in the infrared frame has important implications for the analysis of stellar orbits (see section~\ref{sec:discussion}). To improve this precision we have implemented a more accurate distortion correction for the S27 camera of NACO. In section~\ref{sec:distortion_correction} the apparent distortion is measured at four epochs between 2004 and 2012 from images of globular clusters in the ESO archive. This is possible by matching them to Hubble Space Telescope~(HST) astrometry, which can be considered distortion-free in comparison. Based on the resulting distortion models it is then possible to also correct other images taken close in time. A per night self-calibration approach is not feasible for our Galactic Center observations, because the requirement that many of the same stars are placed in many different regions on the detector is not met. To start with, we describe our data and basic data processing in section~\ref{sec:image_processing}.

\begin{table*}
\centering
\begin{minipage}{165mm}
\caption{Summary of globular cluster observations: A distortion correction for NACO.}
\label{tab:summary1}
\begin{tabular}{llllllllll}
 \hline
  & ESO Prog. ID & Date & Target & Filter & DIT (s) & NDIT & Images & Sample Size & Indiv. Sources \\
  \hline
 (a) & 074.D-0151(A) & 2004-10-12 & 47 Tuc. & Ks & 2.5 & 24 & 20 & 19826 & 1698 \\
 (b) & 482.L-0793(A) & 2009-07-03 & $\omega$ Cen. & Ks & 0.3454 & 100 & 22 & 3910 & 534 \\
 (c) & 60.A-9800(J) & 2010-02-07 & $\omega$ Cen. & Ks & 1. & 10 & 20 & 3658 & 388 \\
 (d) & 089.C-0638(A) & 2012-07-22 & 47 Tuc. & Ks & 1. & 30 & 10 & 9671 & 1824 \\
 \hline
\end{tabular}
\end{minipage}
\end{table*}

\begin{table*}
\centering
\begin{minipage}{165mm}
\caption{Summary of Galactic Center observations: The Sgr~A* rest frame.}
\label{tab:summary2}
\begin{tabular}{llllllllll}
 \hline
 ESO Prog. ID & Date & Filter & DIT (s) & NDIT & Pointings & PSF & Maser & Reference & Distortion \\
  &  &  &  &  & $\times$ Images & FWHM (px) & Stars & Stars & Correction \\
  \hline
 60.A-9026(A) & 2002-04-01 & Ks & 0.5 & 8 & 10$\times$1 & 3.8 & 7 & 73 & model (a) \\
 71.B-0077(A) & 2003-05-09 & Ks & 0.5 & 120 & 19$\times$1 & 2.7 & 8 & 91 & model (a) \\
 073.B-0085(E) & 2005-05-12 & Ks & 0.5 & 60 & 16$\times$6 & 2.9 & 8 & 91 & model (a) \\
 077.B-0014(A) & 2006-04-28 & Ks & 2.0 & 28 & 8$\times$4 & 3.3 & 7 & 87 & model (a) \\
 077.B-0014(E) & 2006-08-27 & Ks & 2.0 & 28 & 8$\times$4 & 4.5 & 7 & 79 & model (a) \\
 078.B-0136(B) & 2007-03-16 & Ks & 2.0 & 28 & 7$\times$4 & 2.6 & 8 & 91 & model (b) \\
 179.B-0261(A) & 2007-03-31 & Ks & 5.0 & 6 & 16$\times$6 & 2.6 & 8 & 91 & model (b) \\
 179.B-0261(M) & 2008-04-04 & Ks & 2.0 & 28 & 8$\times$4 & 2.8 & 8 & 91 & model (b) \\
 179.B-0261(N) & 2008-08-04 & Ks & 1.0 & 57 & 8$\times$4 & 3.0 & 8 & 91 & model (b) \\
 179.B-0261(U) & 2008-09-15 & Ks & 1.0 & 57 & 8$\times$4 & 2.7 & 8 & 91 & model (b) \\
 179.B-0261(X) & 2009-03-28 & Ks & 1.0 & 60 & 8$\times$4 & 2.9 & 8 & 91 & model (b) \\
 179.B-0261(X) & 2009-03-30 & Ks & 1.0 & 60 & 8$\times$4 & 2.7 & 8 & 91 & model (b) \\
 183.B-0100(J) & 2009-09-19 & Ks & 1.0 & 60 & 16$\times$4 & 2.7 & 8 & 91 & model (b) \\
 183.B-0100(J) & 2009-09-20 & Ks & 1.0 & 60 & 8$\times$4 & 2.8 & 8 & 91 & model (b) \\
 183.B-0100(T) & 2010-05-08 & Ks & 0.5 & 126 & 8$\times$4 & 3.2 & 8 & 90 & model (c) \\
 183.B-0100(V) & 2010-09-27 & Ks & 1.0 & 126 & 8$\times$2 & 3.4 & 8 & 86 & model (c) \\
 183.B-0100(X) & 2011-04-01 & Ks & 1.0 & 66 & 8$\times$4 & 3.1 & 8 & 86 & model (c) \\
 183.B-0100(V) & 2011-05-16 & Ks & 2.0 & 9 & 8$\times$2 & 2.6 & 8 & 91 & model (c) \\
 088.B-1038(A) & 2012-03-14 & Ks & 1.0 & 30 & 41$\times$1 & 3.3 & 8 & 86 & model (d) \\
 088.B-0308(B) & 2012-05-03 & Ks & 0.9 & 33 & 8$\times$6 & 3.4 & 8 & 86 & model (d) \\
 089.B-0162(D) & 2012-08-08 & Ks & 1.0 & 60 & 8$\times$4 & 2.7 & 8 & 91 & model (d) \\
 091.B-0081(F) & 2013-05-13 & Ks & 0.9 & 33 & 16$\times$6 & 2.6 & 8 & 90 & model (d) \\
 \hline
\end{tabular}
\end{minipage}
\end{table*}

\section{Image Processing}
\label{sec:image_processing}

\begin{figure} 
\includegraphics[width=84mm]{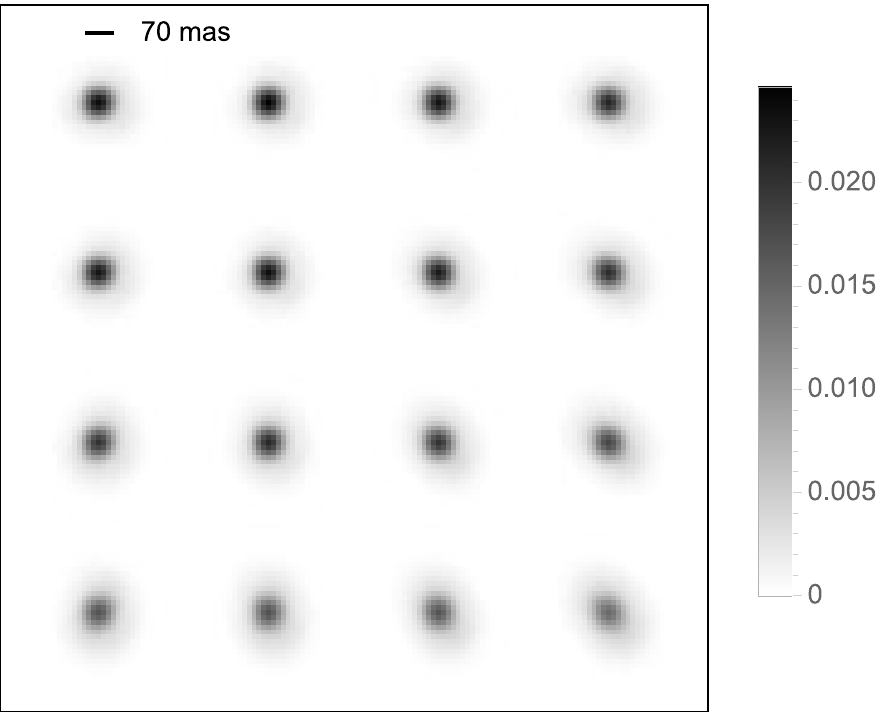}
\caption{Example of a PSF grid obtained from a Galactic Center image with median Strehl ratio. Each PSF is normalized to unit flux. The adaptive optics guide star is located near the top of the image. The typical deviation from the mean PSF in a single pixel is about $10\%$ to $35\%$.}
\label{fig:psf}
\end{figure}

\subsection{Observations}
\label{subsec:observations}

We make use of two sets of observations with NACO/S27 for different purposes, which target the Galactic Center, but also the densely populated cores of the two globular clusters $\omega$~Centauri and 47~Tucanae. The globular cluster observations that result in viable distortion corrections are summarized in table~\ref{tab:summary1}. The instrument was configured to use the Ks-band filter, the Aladdin~3 detector in the high dynamic range mode and the readout mode read-reset-read. This is also the typical setup for many epochs of Galactic Center observations using the same camera, which together establish the astrometric reference frame and are summarized in table~\ref{tab:summary2}. A small number of images are omitted, for example those compromised by severe saturation of the relevant SiO maser stars or those affected by PSF artefacts caused by a failure in the adaptive optics system. The dithering scheme is typically a wide square pattern, such that the total field of view is $42''\times42''$ and the central $14''\times14''$ are present in all images. At each pointing position either two, four or six images are taken, before moving on. A small offset of a few pixels is applied on revisiting a pointing position to minimize the effect of detector artefacts on mosaic images.

We also make use of high-quality HST proper motion catalogues available for $\omega$~Centauri \citep{2010ApJ...710.1032A} and 47~Tucanae (Bellini et al., 2014, private communication), which provide us with a nominally distortion-free reference system. The HST astrometry profits from the stable observing conditions in space and thanks to a well-characterized PSF and distortion correction for the instrument channels utilized, the measurement uncertainty of stellar positions on a single image is as low as ${\sim0.01\rmn{~px}}$ or ${\sim0.5\rmn{~mas}}$ \citep[e.g.][]{2006acs..rept....1A}. More details on the HST observations can be found in \citet{2014ApJ...797..115B} and references therein.

\subsection{Image Reduction}
\label{subsec:image_reduction}

The standard steps of sky subtraction, division of a flat-field and a bad pixel correction are applied to every raw NACO image. The nightly sky background is estimated by taking the median image of either a set of dedicated sky exposures or else the randomly dithered science exposures. The flat-field is created by stacking and normalizing lamp exposures taken during daytime calibrations. The unusable hot or dead pixels are identified by comparing the ADU counts in neighbouring pixels, from which a replacement value is interpolated if necessary. Individual detector integrations were recorded for most images since 2010, so that frames of exceptionally low quality can be rejected before the remaining frames are averaged to create the final image.

\subsection{Star Lists}
\label{subsec:star_lists}

With the aim of achieving the highest astrometric precision, we extract the detector positions (and fluxes) of stars in every reduced NACO image using an empirical model of the PSF \citep[see][]{2000PASP..112.1360A}. To identify the stars, an image is first correlated with the current PSF estimate, with the correlation operator being the normalized cosine distance\footnote{This distance is ${\frac{u-\overline{u}}{|u-\overline{u}|}\cdot\frac{v-\overline{v}}{|v-\overline{v}|}}$ between two vectors $u$ and $v$, which are here one-dimensional arrangements of the correlation kernel and the neighbourhood of a certain pixel, respectively.}. Initially, a Gaussian kernel is substituted for the PSF and since this kernel is symmetric, all PSFs derived from it are naturally centred. The result is a map of coefficients between $-1$ and $1$, which measure at every pixel how well the image matches to the PSF locally. The location of a candidate star is the peak of a connected pixel region in this map, for which the correlation coefficient exceeds a threshold value of $0.7$ and the ADU counts fall in the range between the noise level and the full-well capacity of the detector. At each candidate position, the PSF is fit to a roughly circular, top-hat image region by method of least squares. There are a free position offset and a free scaling factor, as well as three additional parameters that describe a tilted planar background, which minimizes a potential bias in crowded regions. The radius of the fit region, i.e. the number of pixels surrounding a central pixel at all sides, is typically ${4\rmn{~px}}$, but tied to the full width at half-maximum (FWHM) of the PSF such that the entire PSF core is included.

The actual PSF not only changes with time, but varies spatially too. For example, the PSF gets significantly broader and elongated at distances of $10''$ to $20''$ from the adaptive optics guide star, as the correction of atmospheric turbulence degrades due to anisoplanatism. To take into account such a variation we use one ${4\times4}$ grid of PSFs per image, each of which is derived from up to fifty ideally bright and isolated stars within cells of ${256\times256~\rmn{px}}$ size spanning the detector (Fig.~\ref{fig:psf}). By using a bicubic interpolation to resample the surrounding image regions, the selected stars are magnified by a factor $2$ and centred in sub-images typically ${\rmn{35~px}}$ wide, depending on the FWHM of the PSF, and also normalized to zero background and unit flux. The median superposition of these sub-images, weighted by the square root of the original fluxes, is an estimate for the PSF at the centre of a grid cell. Afterwards, a modified cosine window function is used to taper the extended PSF haloes. The PSF at a certain detector position is finally estimated by means of a bilinear inter- or extrapolation on the whole grid, using spline interpolations of third order to evaluate each grid PSF at sub pixel offsets.

The two iterative steps of extracting a PSF grid and creating a star list are repeated three times. In the case of mildly saturated stars, pixels with an ADU count above $80\%$ the full-well capacity are ignored throughout, as long as the fraction of excluded pixels is small (${<10\%}$). Heavily saturated stars are never identified as candidate stars in the first place. The final outcome are lists which contain the positions of all detected stars in the pixel coordinate system of each particular image. They also contain the fluxes of the stars, which are converted into instrumental magnitudes. Only stars brighter than the distribution's peak magnitude are kept (${m_K\lesssim16}$). These are likely to be genuine detections for which positions can be measured with similar precision.

\section{A Distortion Correction for NACO}
\label{sec:distortion_correction}

\begin{figure*} 
\includegraphics[width=84mm]{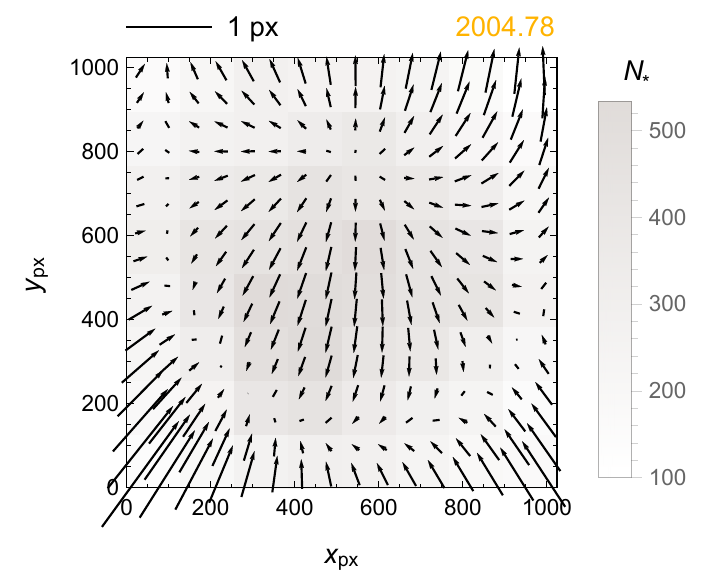}
\includegraphics[width=84mm]{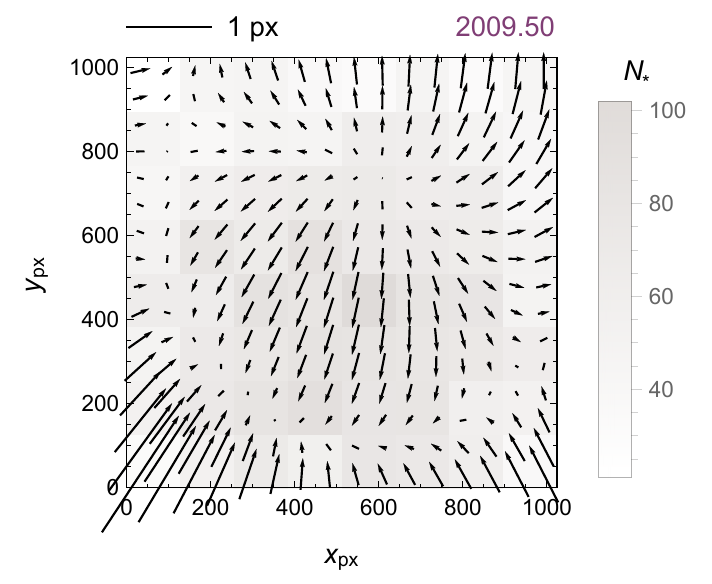}\\
\includegraphics[width=84mm]{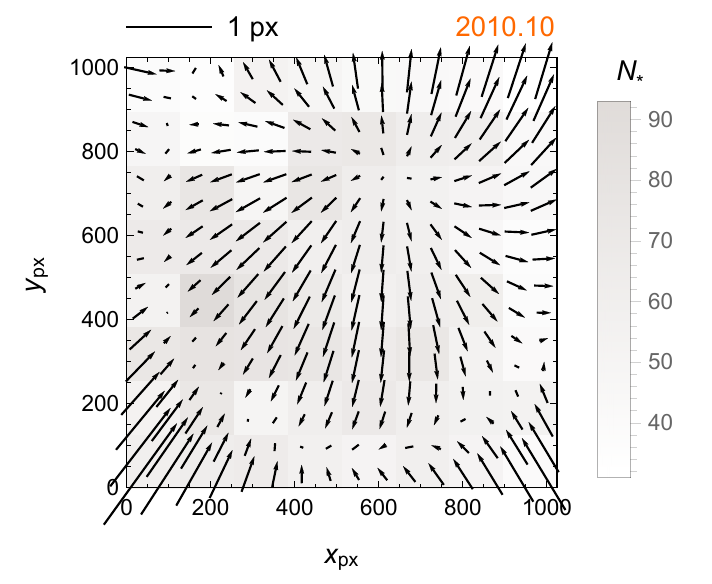}
\includegraphics[width=84mm]{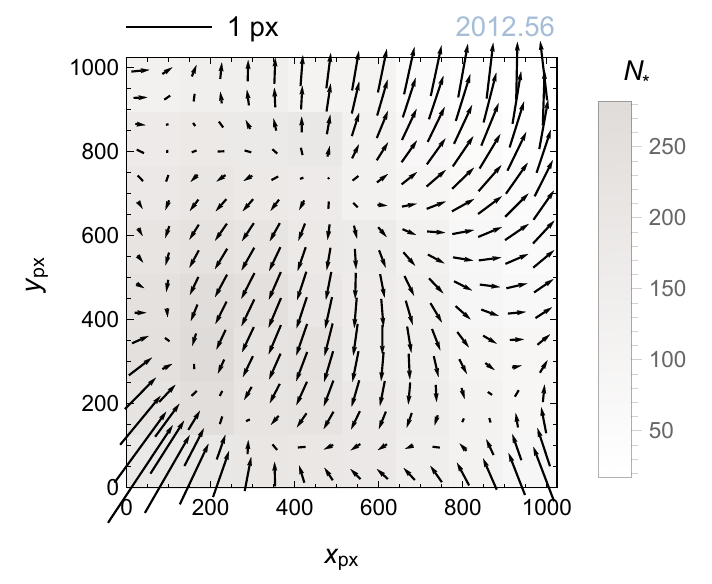}
\caption{The evolution of image distortion in the S27 camera of NACO. Shown is the shift required to move a measured pixel position into a pixel coordinate system approximately free of distortion (black arrows). Also shown is the number of reference sources in different regions on the detector (grey-shaded bins). Each model is only valid for a limited period of time, as shown in Fig.~\ref{fig:timeline}.}
\label{fig:maps}
\end{figure*}

\begin{figure*} 
\includegraphics[width=168mm]{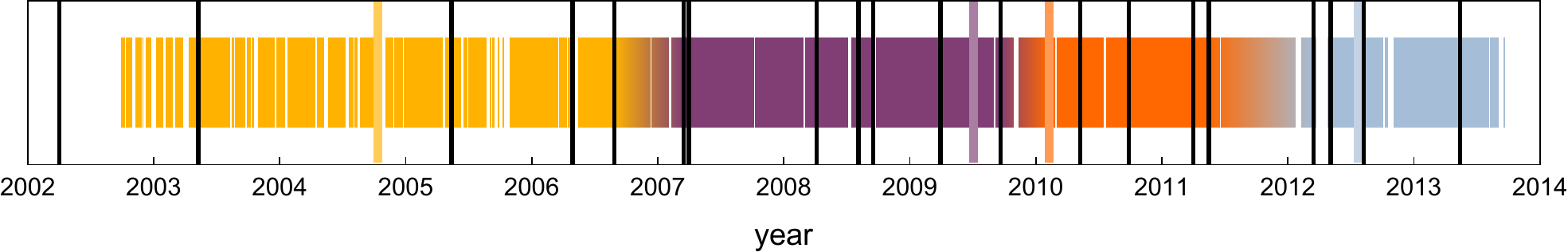}
\caption{Timeline of records in the NACO archive, on which are marked the different epochs of globular cluster observations and the periods of time during which the corresponding distortion corrections are valid (in colour), as well as the epochs of Galactic Center observations that together establish the astrometric reference frame (in black). The exact dates of significant changes in the image distortion might coincide with interventions on the instrument, but cannot be determined exactly from our data. However, it is straightforward to decide whether a certain distortion correction is applicable by comparing the resulting image registration accuracy.}
\label{fig:timeline}
\end{figure*}

\subsection{Star Matching}
\label{subsec:star_matching}

To measure the distortion in the globular cluster images, the respective star lists (Section~\ref{subsec:star_lists}) are first matched to a proper motion catalogue obtained with the HST. A reference star list is created for each observing night by propagating all catalogue positions to the right date. Since the images from a single night overlap, it is convenient to determine the relative offsets between the NACO star lists by cross-correlating the original images. Initial transformations that shift and scale the HST star list to roughly align with each NACO star list are determined manually. Matching pairs of stars are then found using a nearest neighbour search. For each detected star, to make a match with a catalogue star, there has to exist exactly one star within a certain radius around it on the transformed HST list and no other star on its own NACO star list. The latter requirement effectively excludes stars with close neighbours. Additionally, the match needs to be unique. Once matching pairs of stars are known, they can be used to find better transformations for the original HST star list and another matching process follows. The cutoff radius for the nearest neighbour search is set to half the FWHM of the PSF, but five times larger for the first iteration. A last quality criterion is the number of matched stars per image. It should be greater than twenty to allow the combination of multiple star lists later on. Typically, a few thousand detector positions of a few hundred individual stars are accepted in total, which cover the detector reasonably well.

\subsection{The Distortion Model}
\label{subsec:distortion_model}

The image distortion we aim to correct is expected to be static, but might still change sporadically due to differences in the optical alignment before and after interventions on the instrument, for example. All our Galactic Center observations, spread over the whole time-range of NACO operations from 2002 to 2013, are in the end covered by four different distortion corrections (Fig.~\ref{fig:maps} \& Fig.~\ref{fig:timeline}).

At each of the according four epochs of globular cluster observations (Table~\ref{tab:summary1}), a linear transformation is determined for every NACO star list that aligns the matched reference positions from the HST catalogue with the stellar positions as measured on the detector, in a least squares sense. More precisely, it is an affine transformation (with six parameters), which already combines translation, rotation, scaling and shear operations. The map of transformation residuals, i.e. the remaining differences between aligned positions, is a vector field that represents the image distortion. However, to create a single map from multiple images, it is necessary to at least once apply a preliminary distortion correction to the NACO star lists, recalculate the transformations and refit a model to the pooled residuals. This iteration reduces a potential bias that occurs if stellar positions from individual lists are not distributed uniformly over the detector.

The distortion model we adopt is a linear combination of vector fields from a complete orthonormal basis that can describe arbitrary vector fields on the unit circle (Table~\ref{tab:polynomials}). Such a basis can be constructed from the union of a set of vector polynomials with zero curl and another set with zero divergence \citep[see][]{2007OExpr..1518014Z,2008OExpr..16.6586Z}. Both sets are needed to account for aberrations of the optical system and the arrangement of the detector. This model is fit to each of the four distortion maps using a least squares technique, after the detector coordinates have been rescaled accordingly and the maps have been smoothed with a median filter to clean them from outliers (e.g. falsely matched stars). We pick a smoothing length of $64~\rmn{px}$ to roughly correspond to the scale at which the median absolute deviation between the smoothed and the original vector fields becomes approximately constant.

Twenty free parameters are needed to fully capture the spatial variability of the image distortion. This number can be inferred by inspecting the fit residuals, which stop decreasing when enough basis fields of higher order have been included. Even so, most of the distortion can be attributed to the low-order aberrations $y$-tilt, defocus, coma and spherical aberration. In terms of the number of parameters, an equivalent model could be composed of two third-order polynomials in two variables, each of which describes the distortion along one dimension.

The characteristic shift required to move a measured position into a pixel coordinate system approximately free of distortion is ${\sim0.2~\rmn{px}}$ (${\sim5.4~\rmn{mas}}$), but the actual shift varies strongly with location on the detector and can be as high as ${\sim0.7~\rmn{px}}$ (${\sim19~\rmn{mas}}$) in the lower left corner. The distortion pattern changes distinctly as the pairwise rms deviation between subsequent models is on the order of ${\sim0.1\rmn{~px}}$ (${\sim2.7~\rmn{mas}}$), but the general pattern is static.

\section{The Sgr A* Rest Frame}
\label{sec:astrometry}

\subsection{Image Registration}
\label{subsec:image_registration}

To be able to measure the motions of stars later used as astrometric reference sources, individual images from each night of Galactic Center observations (Table~\ref{tab:summary2}) must first be combined into a common coordinate system.

We start by combining the star lists of consecutive images belonging to the same pointing position, which share similar observing conditions. Stars are iteratively matched between those lists, allowing for small offsets between the images (${\lesssim0.2~\rmn{px}}$). New star lists are then created from the mean positions, but, to further purge the lists of spurious detections, only stars detected in all images are kept. The measurement uncertainty of the new positions is estimated by the standard error of the mean and has a typical value of about ${\sim0.3~\rmn{mas}}$ (mode of histogram).

The single-pointing star lists are then combined into one master star list per epoch. Affine transformations should suffice to register the star lists after they have been corrected for image distortion, since any remaining non-linear displacements should be negligible (considering for example atmospheric refraction, aberration due to the motion of the Earth, light deflection in the gravitation field of the Sun or curvature of the celestial sphere). Nevertheless, we explicitly correct the star lists for differential achromatic refraction as well, to minimize anisotropy of the pixel scale in the final master list \citep[see][]{1998PASP..110..738G}.

The initial master star list is the first star list containing the maximum number of SiO maser stars. We proceed by iteratively matching to it the list with the next smallest pointing offset, finding an affine transformation between the matched positions and creating a star catalogue that contains all aligned positions, grouped by star (unmatched positions are simply appended). The new master list is a collapsed version of this catalogue, i.e. it contains the weighted average positions and the mean uncertainties. In this way both the catalogue and the master star list are updated until all lists are merged. The whole stitching process is repeated once, with the only change being that every star list is matched to the same intermediate master star list.

\begin{figure}
\includegraphics[width=84mm]{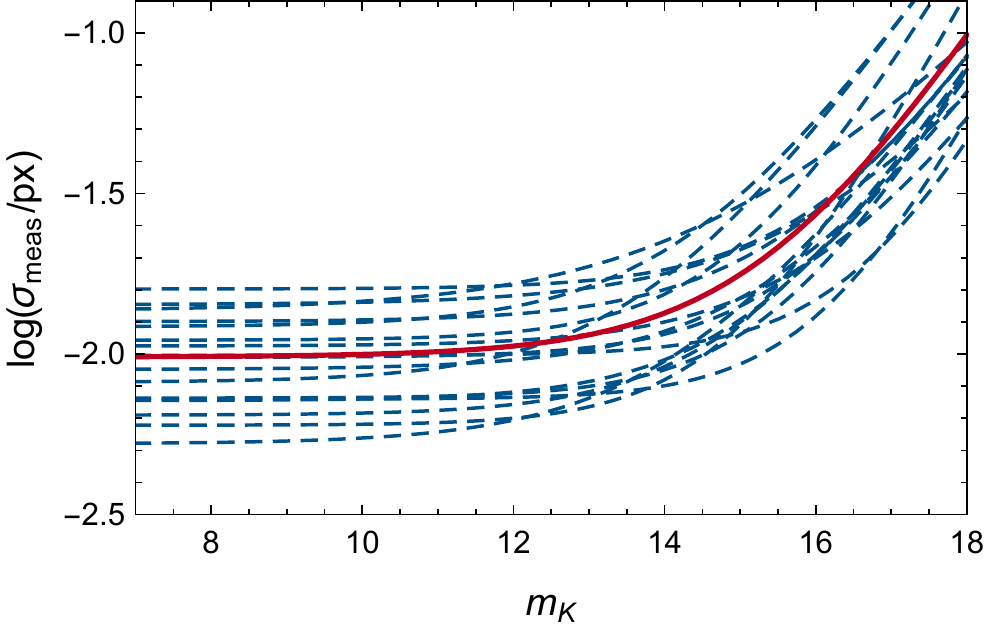}
\caption{Best-fitting models for the measurement uncertainty of stellar positions as a function of K-band magnitude (dashed lines) and the mean model (solid line), which is applied whenever the uncertainty cannot be estimated reliably from the images themselves (see text).}
\label{fig:error_model}
\end{figure}

A few epochs are treated differently in some ways. For observations before 2004, some image reduction parameters have to be adjusted for the different characteristics of the Aladdin~2 detector, which was replaced during that year. For observations with either a random dithering scheme or less than three images per pointing, the measurement uncertainty of stellar positions cannot be estimated reliably from the images themselves. Instead, we apply a mean error model based on the other observations by fitting a power law to the one-dimensional uncertainties of consistently detected stars as a function of flux  (Fig.~\ref{fig:error_model}). The photometric zero-point is calibrated using the $91$ primary astrometric reference stars \citep[section~\ref{subsec:alignment},][]{2009ApJ...692.1075G}.

\begin{figure}
\includegraphics[width=84mm]{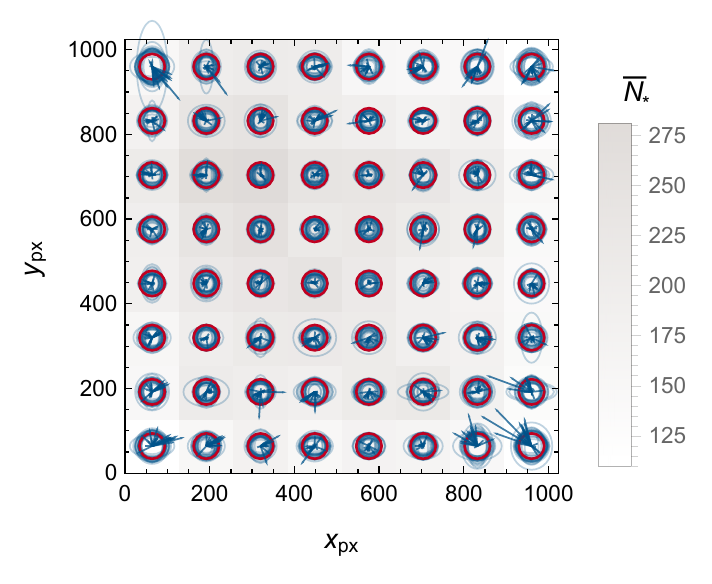}
\caption{The mean (blue arrows) and standard deviation (blue circles) of the standardized registration residuals in bins on the detector, overlaid for all epochs. A systematic deviation from a standard normal distribution (red circles) indicates uncorrected distortion or unrealistic uncertainties.}
\label{fig:residuals}
\end{figure}

\subsection{Testing the Distortion Correction}
\label{subsec:testing}

After the image registration step, any uncorrected distortion becomes apparent in non-random structure of the final transformation residuals. This is because the positions of the same stars, measured in different locations on the detector, should align nevertheless. We can thus check the quality of the distortion corrections and identify approximately the periods during which a particular solution is valid (Fig.~\ref{fig:timeline}).

Every transformation residual is divided by its expected uncertainty, i.e. the combined uncertainty of the two respective positions, and sorted into bins of ${128~\rmn{px}\times128~\rmn{px}}$ size on the detector (Fig.~\ref{fig:residuals}). The uncertainties of the transformation parameters are negligible, given that a few hundred well-measured stars can always be matched between two star lists. An additional statistical uncertainty of ${\sim0.1~\rmn{mas}}$ needs to be added (in quadrature) to each stellar position to make the overall distribution of the so-standardized residuals approximate a standard normal distribution. However, the agreement is best at the centre of the detector and worse at the edges, where the remaining time-averaged distortion is usually at most ${\sim0.2~\rmn{mas}}$. We therefore add a different uncertainty in each bin and then average over a star's detector positions. The distortion model is inherently less constrained at the edges of the detector, but the actual distortion is also expected to vary on a low level even during the night, for example due to unstable performance of the adaptive optics system.

\begin{figure*} 
\includegraphics[width=180mm]{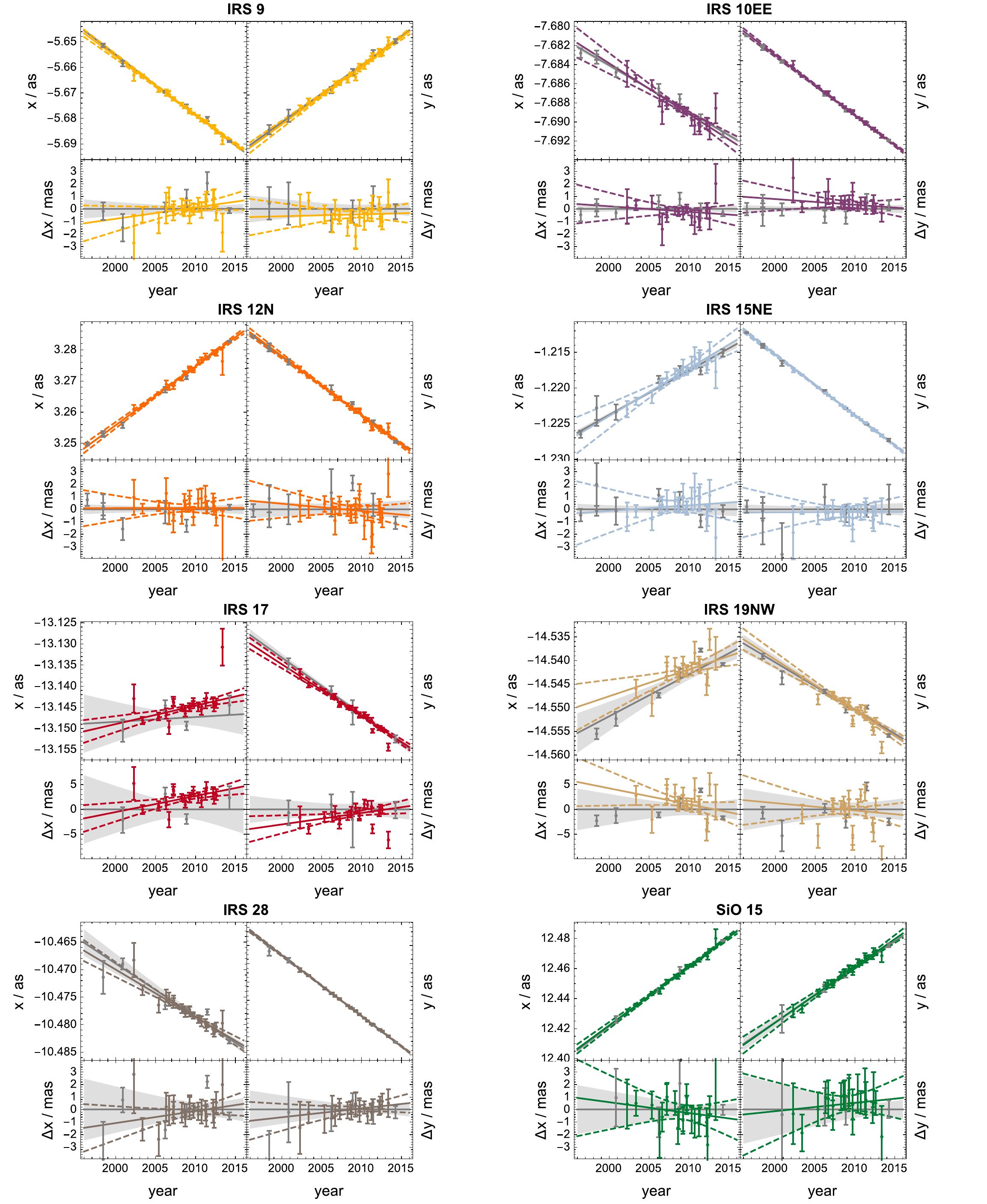}
\caption{The infrared positions of the SiO maser stars and the best-fitting proper motion models (in colour). In each panel the top plots show the motion relative to Sgr~A* and the bottom plots show the motion relative to the radio reference measurements (in grey).}
\label{fig:maser_motion}
\end{figure*}

\begin{figure*}
\includegraphics[width=168mm]{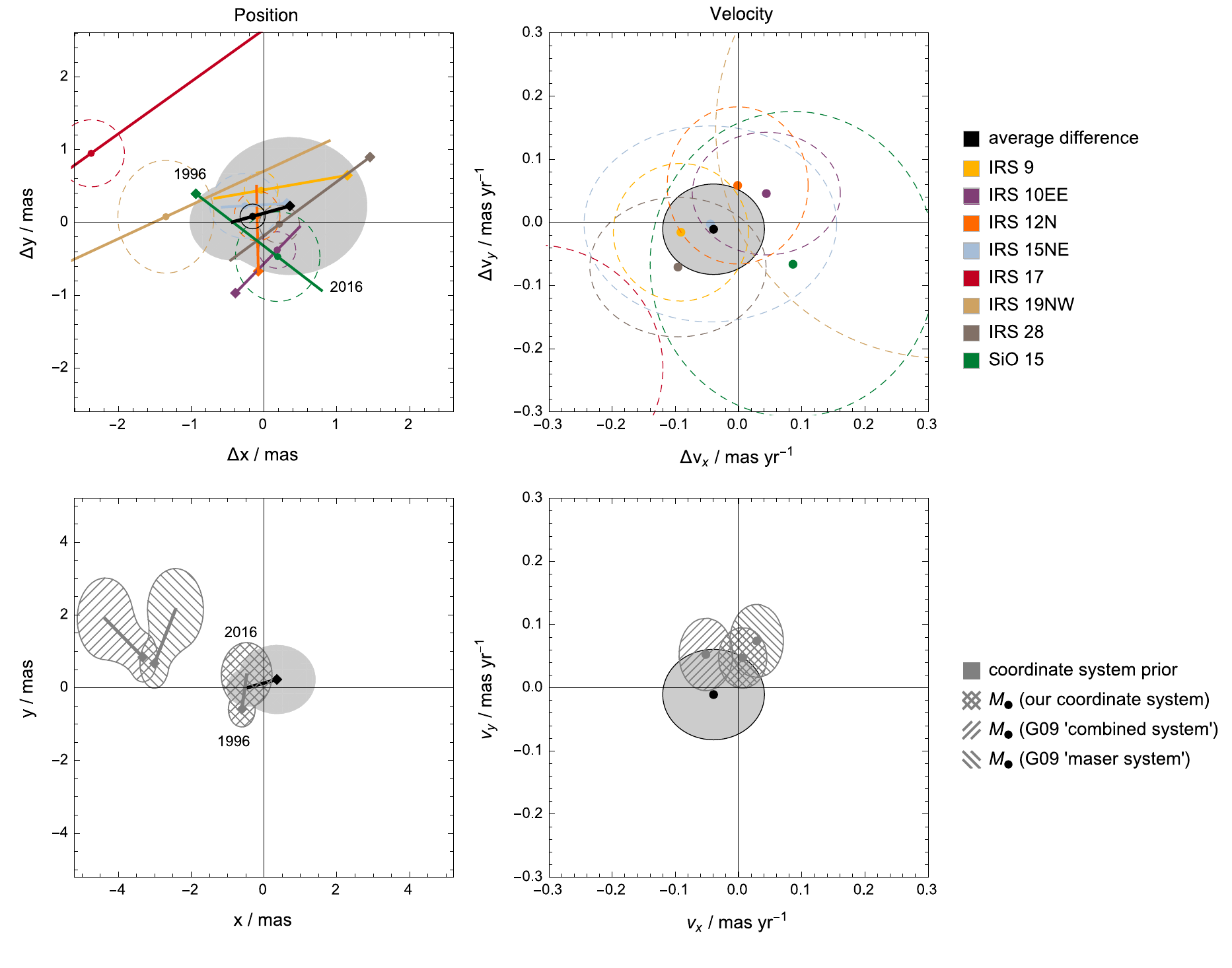}
\caption{Top: Differences between the radio- and infrared-measured linear motions of the SiO maser stars in position (coloured lines on the left) and velocity (coloured dots on the right), shown with their $1\sigma$-uncertainties (dashed ellipses). The position differences are evaluated between 1996 (marked by diamond symbols) and 2016 (unmarked endpoints), with the average reference epoch being $2009.02$ (marked by dots in between). Bottom: The drift motion of the central mass required by the best-fitting orbit of the star S2, using our coordinate system (cross-hatched region) compared with the `combined system' and the `maser system' from \citet[][single-hatched regions]{2009ApJ...692.1075G}. The average deviation shown above (in black and grey) indicates how precisely (radio-)Sgr~A* is localized in the infrared reference frame and is used in the fit as prior information on this drift motion, as indicated here (see section~\ref{subsec:localization}).}
\label{fig:origin_motion}
\end{figure*}

\subsection{The Astrometric Reference Stars}
\label{subsec:alignment}

The proper motions of several late-type giant stars have been measured relative to Sgr~A* directly at radio wavelengths by observing their circumstellar SiO maser emission \citep{1997ApJ...475L.111M,2003ApJ...587..208R,2007ApJ...659..378R}. The results have since been updated with the addition of more data and an improved analysis, including in particular a correction of differential precession, which leads to an apparent rotation (Reid et al., 2015, private communication). Eight of these maser stars lie in the central $7''$ to $24''$ and are typically inside the NACO field of view. Not included is the exceptionally bright star IRS~7, because saturation of the detector prevents an accurate measurement of its infrared position. As the only red supergiant in the sample, its radio position would also have a considerably larger uncertainty than the ${\sim0.5~\rmn{mas}}$ intrinsic to red giants, because it is expected to have a larger SiO maser shell (${\gg4~\rmn{AU}}$) and the variation of the maser emission on time-scales of ${\sim1~\rmn{yr}}$ is not resolved.

At each of the $22$ epochs of Galactic Center observations, the infrared positions of the maser stars taken from the master star list (Section~\ref{subsec:image_registration}) are aligned with the propagated radio positions using a weighted affine transformation, thus taking into account the uncertainties in both sets of positions. The pixel coordinates $(x_\rmn{px},y_\rmn{px})$ of all detected stars are thereby converted into angular offsets $(x,y)$ from Sgr~A*\footnote{$+x$ increases to the west and $+y$ to the north. The conversion to celestial coordinates is given to good approximation by ${(x,y)=(-\Delta\alpha\cos(\delta),\Delta\delta)}$.}. The positional uncertainties, including both measurement uncertainty and the extra uncertainty due to residual distortion (Section~\ref{subsec:testing}), are propagated using a Monte Carlo Bootstrap technique, along with the uncertainties of the transformation parameters themselves.

A total of $10^4$ realizations of one transformation are generated by applying random displacements to the infrared and radio positions of the maser stars, according to their known uncertainties, while simultaneously resampling the pairs of corresponding positions with replacement. The artificial transformations are applied to the other stars as well, whose pixel positions are repeatedly perturbed in the same way. A small number of potential realizations are excluded~(${<1\%}$), since at least three unique pairs of positions are needed to define an affine transformation. The uncertainty of each star's astrometric position is finally estimated by the standard deviation of the sample of transformed positions, which in turn is estimated by the median absolute deviation. This statistic is more robust against outliers than for example the rms deviation. Likewise, the uncertainty of the detector position of Sgr~A*, typically ${\sim0.6\rmn{~mas}}$, can be found by inverting the transformations and finding the pixel positions that map on to the origin. Although the typical uncertainty of the maser stars' detector positions is ${\sim0.3\rmn{~mas}}$, that of the astrometric positions is ${\sim0.8\rmn{~mas}}$ and evidently dominated by uncertainty in the alignment. The main reason is the thinly scattered distribution of the maser stars across the field of view, but also the additional uncertainty of the radio positions.

We fit each maser star's astrometric positions as a function of time with both a linear and a quadratic proper motion model, separately in $x$ and $y$ (Fig.~\ref{fig:maser_motion} \& Table~\ref{tab:maser_motion}). For simplicity we use one reference epoch per star instead of one per fit and thus correlations between fit parameters are not entirely eliminated. The uncertainties of the fit parameters are estimated from a Monte Carlo sample created by performing fits to the many positions of each maser star generated at every epoch, which are additionally resampled with replacement in time. The mean reduced $\chi^2$ of $0.97$ suggests that the positional uncertainties have possibly been slightly overestimated.

Whether a detected acceleration is genuine is decided by two criteria. First, the direction of the acceleration must be towards Sgr~A*, i.e. the negative radial component must be statistically significant while the tangential component must be insignificant, at the $5\sigma$ level. Second, the magnitude of the acceleration must be smaller than the upper limit imposed by the gravitational force the massive black hole associated with Sgr~A* can exert, assuming that this is the dominant force and given that the observed projected separation from Sgr~A* is the minimum true separation. None of the best-fitting quadratic models satisfy both criteria and we keep the linear models.

Another step of calibration is necessary to measure the astrometric positions of the S-stars on images taken with the S13 camera, because the maser stars are either saturated in the deeper exposures or outside the smaller field of view. The connection between the two image scales is made by a sample of $91$ astrometric reference stars, which can always be observed together with either the S-stars (using NACO/S13) or the maser stars (using NACO/S27). These reference stars lie in the central $0.8''$ to $5''$ and were chosen to be relatively isolated from other known sources \citep{2009ApJ...692.1075G}. We fit proper motion models for the reference stars analogously to the maser stars. Since also none of them show significant plausible accelerations, we discard the quadratic models. Lastly, we inspect the fit residuals separately for individual stars and epochs. Particularly in 2013 there is a systematic pattern originating from a misalignment of the maser stars, but it is consistent with the statistical uncertainties.

The overall detection rate of both the reference and the maser stars is about $97\%$. Some fainter reference stars were sometimes not detected in a pointing when they could have been, due to a combination of varying image quality and confusion. Non-detections of the maser stars were caused by saturation of IRS~17 and IRS~9, as well as the placement of IRS~12N close to the edge of the north-east images.

\begin{figure}
\includegraphics[width=84mm]{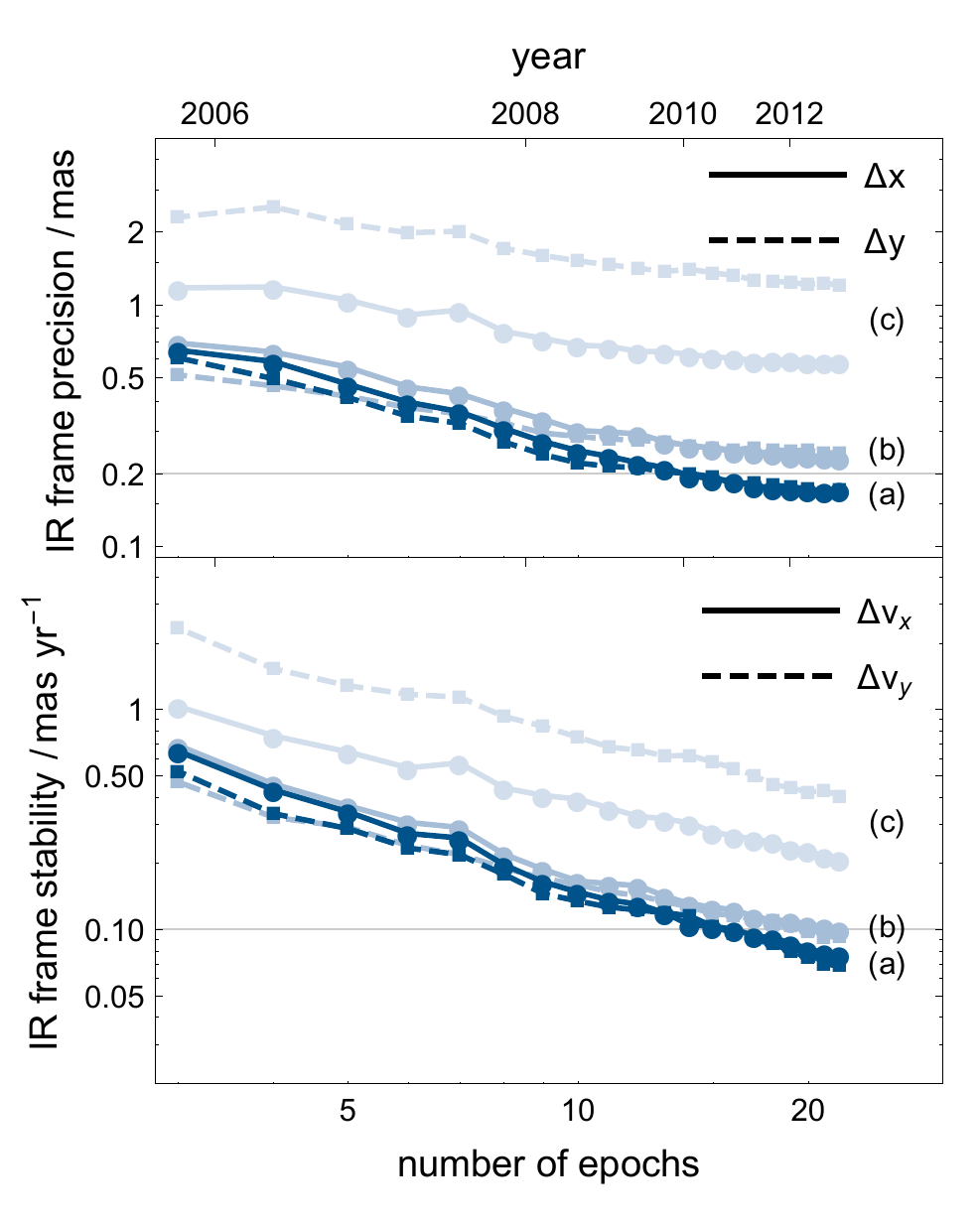}
\caption{Precision and stability of the Sgr~A* rest frame as a function of time when excluding past infrared observations (a) and also either excluding radio observations after 2010 (b) or not applying a distortion correction (c).}
\label{fig:frame_prop}
\end{figure}

\begin{table*}
\centering
\begin{minipage}{140mm}
\caption{Proper motions of the SiO maser stars in the Sgr~A* rest frame.}
\label{tab:maser_motion}
\begin{tabular}{lccccc}
 \hline
 ID & $t_0~\rmn{(year)}$ & $x_0~\rmn{(mas)}$ & $y_0~\rmn{(mas)}$ & $v_x~(\rmn{mas~yr}^{-1})$ & $v_y~(\rmn{mas~yr}^{-1})$ \\
 \hline
IRS 9 & 2009.17 & -5676.73 $\pm$ 0.23 & -6333.35 $\pm$ 0.26 & -2.29 $\pm$ 0.11 & 1.62 $\pm$ 0.11 \\
IRS 10EE & 2009.00 & -7688.65 $\pm$ 0.26 & 4186.87 $\pm$ 0.25 & -0.53 $\pm$ 0.12 & -2.87 $\pm$ 0.10 \\
IRS 12N & 2008.56 & 3271.96 $\pm$ 0.31 & -6912.22 $\pm$ 0.34 & 1.88 $\pm$ 0.11 & -2.49 $\pm$ 0.12 \\
IRS 15NE & 2008.43 & -1218.31 $\pm$ 0.46 & 11260.96 $\pm$ 0.43 & 0.67 $\pm$ 0.20 & -5.90 $\pm$ 0.16 \\
IRS 17 & 2009.04 & -13145.03 $\pm$ 0.46 & 5542.86 $\pm$ 0.46 & 0.44 $\pm$ 0.20 & -2.35 $\pm$ 0.20 \\
IRS 19NW & 2009.55 & -14542.15 $\pm$ 0.63 & -18483.37 $\pm$ 0.75 & 0.57 $\pm$ 0.36 & -2.38 $\pm$ 0.36 \\
IRS 28 & 2009.44 & -10478.15 $\pm$ 0.27 & -5853.57 $\pm$ 0.26 & -0.86 $\pm$ 0.14 & -6.64 $\pm$ 0.11 \\
SiO 15 & 2009.28 & 12458.47 $\pm$ 0.58 & -11060.20 $\pm$ 0.61 & 3.90 $\pm$ 0.23 & 2.15 $\pm$ 0.24 \\
 \hline
\end{tabular}
\end{minipage}
\end{table*}

\begin{table*}
\centering
\begin{minipage}{140mm}
\caption{Alignment of the SiO maser stars (positions are compared at epoch $2009.02$).}
\label{tab:origin_motion}
\begin{tabular}{lcccc}
 \hline
 ID & $\Delta x_0~\rmn{(mas)}$ & $\Delta y_0~\rmn{(mas)}$ & $\Delta v_x~(\rmn{mas~yr}^{-1})$ & $\Delta v_y~(\rmn{mas~yr}^{-1})$ \\
 \hline
IRS 9 & -0.04 $\pm$ 0.23 & 0.44 $\pm$ 0.26 & -0.09 $\pm$ 0.11 & -0.02 $\pm$ 0.11 \\
IRS 10EE & 0.18 $\pm$ 0.26 & -0.38 $\pm$ 0.25 & 0.04 $\pm$ 0.12 & 0.05 $\pm$ 0.10 \\
IRS 12N & -0.09 $\pm$ 0.31 & 0.09 $\pm$ 0.34 & -0.00 $\pm$ 0.11 & 0.06 $\pm$ 0.12 \\
IRS 15NE & -0.26 $\pm$ 0.48 & 0.23 $\pm$ 0.44 & -0.04 $\pm$ 0.20 & -0.00 $\pm$ 0.16 \\
IRS 17 & -2.37 $\pm$ 0.46 & 0.95 $\pm$ 0.46 & -0.32 $\pm$ 0.20 & -0.23 $\pm$ 0.20 \\
IRS 19NW & -1.35 $\pm$ 0.66 & 0.08 $\pm$ 0.78 & 0.32 $\pm$ 0.36 & 0.15 $\pm$ 0.36 \\
IRS 28 & 0.21 $\pm$ 0.28 & -0.03 $\pm$ 0.26 & -0.10 $\pm$ 0.14 & -0.07 $\pm$ 0.11 \\
SiO 15 & 0.19 $\pm$ 0.58 & -0.47 $\pm$ 0.62 & 0.09 $\pm$ 0.23 & -0.07 $\pm$ 0.24 \\
  & & & & \\ 
Average & -0.16 $\pm$ 0.17 & 0.08 $\pm$ 0.17 & -0.04 $\pm$ 0.08 & -0.01 $\pm$ 0.07 \\
 \hline
\end{tabular}
\end{minipage}
\end{table*}

\subsection{The Location of Sgr A*}
\label{subsec:localization}

The best-fitting proper motion models for the SiO maser stars can be compared with the predicted radio motions to assess the precision and stability of the Sgr~A* rest frame over time (Fig.~\ref{fig:origin_motion} \& Table~\ref{tab:origin_motion}). We find that the radio- and infrared-measured motions of the maser stars are consistent with each other and that the average difference motion in units of mas is:
\[
\Delta x(t)\approx(-0.16\pm0.17)+(-0.04\pm0.08)(t-2009.02)
\]
\[
\Delta y(t)\approx(+0.08\pm0.17)+(-0.01\pm0.07)(t-2009.02)
\]
We thus conclude that (radio-)Sgr~A*, i.e. the origin of the radio coordinate system, is localized in the infrared reference frame with a precision of ${\sim0.17\rmn{~mas}}$ in position (in~2009) and ${\sim0.07\rmn{~mas~yr}^{-1}}$ in velocity (${\sim2.7\rmn{~km~s}^{-1}}$). Moreover, the average velocity differences in radial and tangential direction indicate that the infrared reference frame shows neither pumping ($v_r/r$) nor rotation ($v_\phi/r$) relative to the radio system to within ${\sim7.5~\mu\rmn{as~yr}^{-1}\rmn{~as}^{-1}}$ and ${\sim7.0~\mu\rmn{as~yr}^{-1}\rmn{~as}^{-1}}$, respectively.

To calculate the average of the position and velocity differences we have used as weights the uncertainties of the infrared quantities, which implicitly contain the propagated uncertainties of the radio motions. To estimate the uncertainties of the average values we have reused the Monte Carlo sample of proper motion fits. The epoch $2009.02$ at which positions are compared is the average fit reference epoch, with the weights being the uncertainties of the absolute velocities. Because we have also used weighted alignment transformations to begin with, the absolute average values are expected to be nonzero.

For comparison, \citet{2009ApJ...692.1075G} were able to localize Sgr~A* to within ${\sim1.80\rmn{~mas}}$ and ${\sim0.33\rmn{~mas~yr}^{-1}}$ in their `maser system', applying in particular a distortion correction for the S27 camera that only allowed for barrel or pincushion distortion and a free optical axis \citep[3~parameters, see][]{2008A&A...492..419T}. However, by using a more complex `combined system' that involved a velocity calibration in the cluster rest frame (Section~\ref{sec:introduction}), the latter value reduced to ${\sim0.06\rmn{~mas~yr}^{-1}}$. \citet{2010ApJ...725..331Y} were able to localize Sgr~A* to within ${\sim0.57\rmn{~mas}}$ and ${\sim0.09\rmn{~mas~yr}^{-1}}$ in their coordinate system, which is also solely based on maser stars and constructed from independent observations with the NIRC2 imager at the Keck observatory.

The largest deviations between infrared and radio measurements occur for the maser stars IRS~17, IRS~19NW and SiO~15. In the radio, IRS~17 has the least certain proper motion and was only observed at four of eight epochs, as was SiO~15. In the infrared, IRS~19NW is the faintest maser star and its bright neighbour IRS~19 might bias its position. It is also the star farthest from the adaptive optics guide star IRS~7, with the next closer one being SiO~15.

The fact that the uncertainty of the infrared location of Sgr~A* is well defined can be exploited in the analysis of stellar orbits, by setting priors on the allowed drift motion of the central mass\footnote{Another prior at hand is the radial velocity of Sgr~A*.}. The star S2 is currently the most important one for constraining the gravitational potential and the distance to the Galactic Center. By following \citet{2009ApJ...692.1075G} in fitting a Keplerian orbit to the available astrometric and spectroscopic data of S2 and using our updated proper motion models for the astrometric reference stars, we were able to reduce this drift motion significantly (Fig.~\ref{fig:origin_motion}), without needing a cross-calibration to the cluster rest frame.

\section{Discussion \& Conclusions}
\label{sec:discussion}

The radio source and massive black hole Sgr~A* at the Galactic Center can now be placed in the origin of an infrared astrometric reference frame with a precision of ${\sim0.17\rmn{~mas}}$ in position (in~2009) and ${\sim0.07\rmn{~mas~yr}^{-1}}$ in velocity, by aligning the positions of SiO maser stars measured at both infrared and radio wavelengths.

Besides the new data collected in recent years, the factor $5$ improvement over the comparable `maser system' from \citet{2009ApJ...692.1075G} is mainly the result of a better, $20$-parameter correction for optical distortion in the S27 camera of NACO, which we derived by comparing archive images of globular clusters to astrometric reference catalogues obtained with the HST. The shifts of stellar positions on the detector due to image distortion are on the level of a few tenths of a pixel (several mas) and the general distortion pattern is complex but static. The origin of the observed minor, yet distinct and apparently abrupt changes is likely a difference in the optical alignment before and after interventions on the instrument. We make our four individual distortion corrections publicly available\footnote{\url{http://www.mpe.mpg.de/ir/gc/distortion}} with the intention that astrometric studies with a different scientific focus will benefit.

The precision and stability of the Sgr~A* rest frame will continue to improve steadily, under the condition that future infrared and radio observations yield more positions of previously observed or newly identified maser stars (Fig.~\ref{fig:frame_prop}). A fundamental limit arises, should the maser emission not track the centroid of the photospheric emission, owing to the spatial distribution of maser spots in the extended stellar envelopes. Nevertheless, the already gained improvement of the infrared reference frame enables a reanalysis of the S-star astrometry, to be presented in a forthcoming paper (Plewa et al., in preparation).

Precision astrometry at the Galactic Center holds great scientific potential and will eventually lead to the detection of relativistic effects on stellar orbits. Given an orbit with eccentricity $e$ around a black hole of mass $M_\bullet$ at a distance $R_0$, the dominant post-Newtonian effect with an impact on astrometry is Schwarzschild precession. This effect causes an apparent apocentre shift per revolution of \citep{1972gcpa.book.....W}:
\[
\Delta s\approx\frac{6\pi G}{c^2}\frac{M_\bullet}{R_0(1-e)}
\]
An extended mass distribution would cause a counteracting Newtonian precession, but is still very uncertain \citep[e.g.][]{2010RvMP...82.3121G}. In the case of the star S2, a drift of the coordinate system of ${\sim 0.05\rmn{~mas~yr}^{-1}}$ would amount to the magnitude of the former effect (${\Delta s\sim0.8\rmn{~mas}}$) after one orbital period. The ability to detect a relativistic precession of the orbit of S2 therefore hinges on a very stable reference frame, but also depends crucially on the astrometric data obtained at and around the time of the next pericentre passage in 2018.

\section*{Acknowledgements}
We are grateful to Jay Anderson and Andrea Bellini for helpful discussions and for contributing HST data that proved to be extremely valuable to our calibration efforts. We also thank the anonymous referee for comments to improve the paper.


\newpage
\appendix
\section{The Distortion Model}

\begin{table}
\centering
\vspace{1cm}
\caption{Explicit form of the distortion model in terms of its basis vector fields. For a derivation see \citet{2007OExpr..1518014Z,2008OExpr..16.6586Z}.}
\label{tab:polynomials}
\begin{tabular}{lll}
 \hline
  $V$ & $V_x(x,y)$ & $V_y(x,y)$ \\
  \hline
 $S_2$ & $a_2$ & $0$ \\
 $S_3$ & $0$ & $a_3$ \\
 $S_4$ & $\sqrt{2} a_4 x$ & $\sqrt{2} a_4 y$ \\
 $S_5$ & $\sqrt{2} a_5 y$ & $\sqrt{2} a_5 x$ \\
 $S_6$ & $\sqrt{2} a_6 x$ & $-\sqrt{2} a_6 y$ \\
 $S_7$ & $\sqrt{6} a_7 x y$ & $\sqrt{\frac{3}{2}} a_7 (x^2+3y^2-1)$ \\
 $S_8$ & $\sqrt{\frac{3}{2}} a_8 (3x^2+y^2-1)$ & $\sqrt{6} a_8 x y$ \\
 $S_9$ & $2 \sqrt{3} a_9 x y$ & $\sqrt{3} a_9 (x-y) (x+y)$ \\
 $S_{10}$ & $\sqrt{3} a_{10} (x-y) (x+y)$ & $-2 \sqrt{3} a_{10} x y$ \\
 $S_{11}$ & $2 a_{11} x (3 x^2+3 y^2-2)$ & $2 a_{11} y (3 x^2+3 y^2-2)$ \\
 $S_{12}$ & $2 \sqrt{2} a_{12} x (2 x^2-1)$ & $2 \sqrt{2} a_{12} y (1-2 y^2)$ \\
 $S_{13}$ & $2 \sqrt{2} a_{13} y (3 x^2+y^2-1)$ & $2 \sqrt{2} a_{13} x (x^2+3 y^2-1)$ \\
 $S_{14}$ & $2 a_{14} (x^3-3 x y^2)$ & $2 a_{14} y (y^2-3 x^2)$ \\
 $S_{15}$ & $-2 a_{15} y (y^2-3 x^2)$ & $2 a_{15} (x^3-3 x y^2)$ \\
 $T_4$ & $\sqrt{2} b_4 y$ & $-\sqrt{2} b_4 x$ \\
 $T_7$ & $\sqrt{\frac{3}{2}} b_7 (x^2+3 y^2-1)$ & $-\sqrt{6} b_7 x y$ \\
 $T_8$ & $\sqrt{6} b_8 x y$ & $-\sqrt{\frac{3}{2}} b_8 (3 x^2+y^2-1)$ \\
 $T_{11}$ & $2 b_{11} y (3 x^2+3 y^2-2)$ & $-2 b_{11} x (3 x^2+3 y^2-2)$ \\
 $T_{12}$ & $2 \sqrt{2} b_{12} y (1-2 y^2)$ & $2 \sqrt{2} b_{12} x (1-2 x^2)$ \\
 $T_{13}$ & $2 \sqrt{2} b_{13} x (x^2+3 y^2-1)$ & $-2 \sqrt{2} b_{13} y (3 x^2+y^2-1)$ \\
 \hline
\end{tabular}
\vspace{-11cm}
\end{table}

\label{lastpage}

\end{document}